%% file: mn2_GvB_SG.tex
\documentclass[useAMS,usenatbib]{mn2e}
\usepackage{graphicx}
\usepackage{lscape}

\usepackage{times}

\input{journal_defs.tex}

\title[Supergiant Temperatures and Linear Radii\\from Near-Infrared Interferometry]{Supergiant Temperatures and Linear Radii\\from Near-Infrared Interferometry}
\author[G. T. van Belle, M. J. Creech-Eakman and A. Hart]{G. T. van Belle$^{1}$\thanks{E-mail:
gvanbell@eso.org (GvB); mce@inanna.nmt.edu (MCE); aamos@du.edu (AH)}, M. J. Creech-Eakman$^{2}$ and A. Hart$^{3}$\footnotemark[1]\thanks{Contact GvB for reprints.}\\
$^{1}$European Southern Observatory, 85478 Garching, Germany\\
$^{2}$New Mexico Institute of Mining and Technology, Socorro, New Mexico, USA\\
$^{3}$University of Denver, Denver, Colorado, USA}
\begin{document}

\date{Accepted TBD. Received TBD; in original form 2008 September 2}

\pagerange{\pageref{firstpage}--\pageref{lastpage}} \pubyear{2008}

\maketitle

\label{firstpage}

\begin{abstract}
We present angular diameters for 42 luminosity class I stars and 32
luminosity class II stars that have been interferometrically determined with the
Palomar Testbed Interferometer.  Derived values of radius and effective
temperature are established for these objects, and an empirical calibration of
these parameters for supergiants will be presented as a functions of spectral
type and colors.  For the effective temperature versus $(V-K)_0$ color, we find an
empirical calibration with a median deviation of $\Delta T = 70$K in the range of
$0.7 < (V-K)_0 < 5.1$ for LC I stars; for LC II, the median deviation is $\Delta T =
120$K from $0.4 < (V-K)_0 < 4.3$.  Effective temperature as a function of spectral type is also calibrated from these data, but shows significantly more scatter than the $T_{\rm EFF}$ versus $(V-K)_0$ relationship.  No deviation of $T_{\rm EFF}$ versus spectral type is seen for these high luminosity objects relative to luminosity class II giants. Directly determined diameters range up to $400 R_\odot$,
though are limited by poor distance determinations, which dominate the error
estimates.  These temperature and radii measures reflect a direct calibration of these parameters for supergiants from empirical means.
\end{abstract}

\begin{keywords}
infrared: stars, stars: fundamental parameters (radii, temperatures),
techniques: interferometric, stars: supergiants.
\end{keywords}

\section{Introduction}

Supergiants are the signposts that call out the extremes of the land of stellar existence in many regards - mass, composition, nucleosynthesis, linear size, temperature, and mass loss.  As probes of the Eddington limit \citep{ed21}, supergiants help define the nature of hydrostatic equilibrium in stellar atmospheres. Over the past century, establishment of fundamental stellar parameters for these objects has remained a regime dominated by spectroscopic characterization (see, for example, \citet{1973ARA&A..11...29M}).  While our understanding of the atomic, molecular, quantum mechanical, and radiative transfer physics that governs such characterization has attained a degree of magnificently exquisite precision, it remains a characterization of macroscopic parameters through a fundamentally microscopic means.

The advances of stellar interferometry over the past two decades have greatly enabled independent investigations of the macroscopic stellar fundamental parameters - particularly temperature and linear radius - from quantification of the stars on macroscopic scales; namely, through measurement of their angular sizes.  A review of the CHARM2 catalog \citep{2005A&A...431..773R} reveals that angular radius measurements of 58 Luminosity Class I (supergiants) and 38 Luminosity Class II (bright giants) have been made to date.  However these data are contained in 35 unique studies in a variety of wavelengths and with inhomogeneous calibrations and analysis techniques.

In contrast to that menagerie of studies, we present a single data set from the Palomar Testbed Interferometer (PTI) specifically aimed at the characterization of supergiant effective temperatures and linear radii.  PTI has previously been employed to establish such parameters for giant stars \citep{1999AJ....117..521V} and Mira variables \citep{Thompson2002ApJ...577..447T,Thompson2002ApJ...570..373T}, and has shown itself to be uniquely capable in conducting large surveys of stellar angular size.
This paper provides the largest homogeneous data set of this kind and doubles the library of angular size data on giants and supergiants.

\section{Observations}

PTI is an 85 to 110 m baseline H- and K-band (1.6 $\mu$m and 2.2 $\mu$m)
interferometer located at Palomar Observatory in San Diego County, California, and is described in
detail in \citet{col99}.  It has three 40-cm apertures used in pairwise
combination for measurement of stellar fringe visibility on sources that
range in angular size from 0.05 to 5.0 milliarcseconds, being able to resolve individual
sources with angular sizes $\theta > 1.0$ mas in size.
PTI has been in nightly operation since 1997, with minimum downtime throughout the intervening years.
The data from PTI considered herein covers the range from the beginning of 1998 (when
the standardized data collection and pipeline reduction went into place) until the beginning of 2008 (when
the analysis of the data presented in this manuscript was begun).  In addition to the supergiants discussed herein,
appropriate calibration sources were observed as well and can be found {\it en masse} in
\citet{2008ApJS..176..276V}.

The calibration of the supergiant visibility $(V^2)$ data is performed by
estimating the interferometer system visibility ($V^2_{\textrm{\tiny
SYS}}$) using the calibration sources with model angular diameters
and then normalizing the raw supergiant visibility by
$V^2_{\textrm{\tiny SYS}}$ to estimate the $V^2$ measured by an
ideal interferometer at that epoch
\citep{1991AJ....101.2207M,1998SPIE.3350..872B,2005PASP..117.1263V}.
Uncertainties in the system visibility and the calibrated target
visibility are inferred from internal scatter among the data in an
observation using standard error-propagation calculations
\citep{1999PASP..111..111C}. Calibrating our point-like calibration
objects against each other produced no evidence of systematics, with
all objects delivering reduced $V^2 = 1$.

PTI's limiting night-to-night
measurement error is $\sigma_{V^2_{\textrm{\tiny SYS}}}\approx 1.5
-1.8$\%, the source of which is most likely a combination of
effects: uncharacterized atmospheric seeing (in particular,
scintillation), detector noise, and other instrumental effects. This
measurement error limit is an empirically established floor from the
previous study of \citet{bod99}.

From
the relationship between visibility and uniform disk angular size $(\theta_{\rm UD})$, $V^2 = [2 J_1(x) / x]^2$, where $J_1$ is the first Bessel function and spatial frequency $x = \pi B \theta_{UD} \lambda^{-1}$, we may establish uniform disk
angular sizes for the supergiants observed by PTI since the accompanying parameters (projected telescope-to-telescope separation, or baseline, $B$ and wavelength of observation $\lambda$) are well-characterized during the observation.
This uniform disk angular size will be connected to a more physical limb darkened angular size $(\theta_{\rm LD})$ in \S \ref{sec_LD}.

\section{Supporting Data}
\subsection{Spectral Type\label{sec_sptype}}

For consistency of spectral typing, we attempted to resolve spectral type and luminosity classification
of our targets
using typing from Morgan, Keenan, and their co-workers \citep{1953ApJ...117..313J,1973ARA&A..11...29M,1989ApJS...71..245K,2006yCat.3150....0K}.  However, in a
few cases, such typing was not available, and alternate sources were used.

\subsection{Spectral Energy Distribution Fitting\label{sec_sedFitting}}

For calibrator stars and the science targets observed in this investigation, a
spectral energy distribution (SED) fit was performed.  This fit was
accomplished using photometry available in the literature as the
input values, with template spectra appropriate for the spectral
types indicated for the stars in question. The template spectra,
from \citet{pic98}, were adjusted by the fitting routine to account
for overall flux level, wavelength-dependent reddening, and
an initial angular size estimate;
effective temperature was fixed for each of the library spectra based on the
spectral type and luminosity class of the spectra.

Reddening corrections were based upon the empirical
reddening determination described by \citet{1989ApJ...345..245C}, which differs
little from van de Hulst's theoretical reddening curve number 15
\citep{joh68,dyc96}. Both narrowband and wideband photometry in the
0.3 $\mu$m to 30 $\mu$m were used as available, including Johnson
$UBV$ (see, for example, \citet{1963AJ.....68..483E,1972ApJ...175..787E,1971A&A....12..442M}),
Stromgren $ubvy\beta$
\citep{1976HelR....1.....P}, 2Mass $JHK_s$ \citep{2003yCat.2246.....C}, Geneva \citep{1976A&AS...26..275R},
Vilnius $UPXYZS$ \citep{1972VilOB..34....3Z}, $WBVR$ \citep{1991TrSht..63....4K}, and IRAS 12 $\mu$m flux \citep{1984ApJ...278L...1N}; flux calibrations
were based upon the values given in \citet{cox00}.  Magnitudes at 12 $\mu$m were established from the IRAS flux
values using the relationship found in \citet{1995AJ....110.2910H}.

\begin{figure*}
\includegraphics[scale=.66,angle=270]{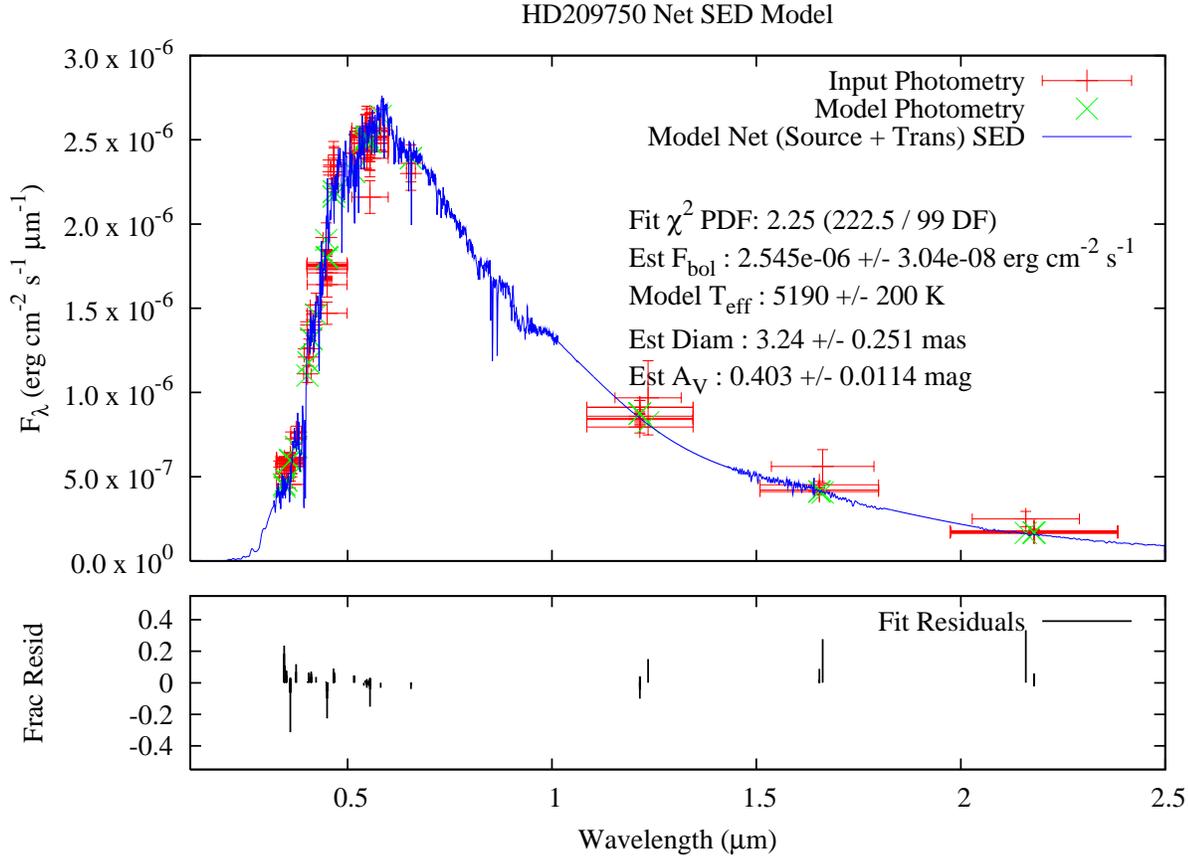}
\caption{\label{fig_HD209750} Spectral energy distribution fitting for HD 209750,
as discussed in \S \ref{sec_sedFitting}, with a G2 I spectral template \citep{pic98} being
fit to the wide- and narrow-band photometry available for the star.
Vertical bars are errors associated with
the photometric data; horizontal bars represent the bandwidth of each
photometric data point.}
\end{figure*}

From the spectral type and luminosity class as discussed in \S \ref{sec_sptype},
template spectra were fitted to the photometric data.
The best SED fit thus provided, for each star, estimates of the bolometric
flux ($F_{\rm BOL}$), angular size ($\theta_{\rm EST}$), and reddening (expressed as magnitudes of visiual reddening $A_{\rm V}$).
The results of
the fitting are given in Tables 6 and 7,
and an example
SED fitting plot is given in Figure~\ref{fig_HD209750}.

\subsection{Distance}

Given the poor relative accuracy of the distances in the Hipparcos catalog \citep{1997A&A...323L..49P}
(unsurprising given the extreme distance of these objects relative to the majority of Hipparcos objects), we opted to use the estimates of \citet{2005A&A...430..165F}
for those objects found in that analysis, where available, reverting to the Hipparcos values
for those stars not found in \citet{2005A&A...430..165F}.

\subsection{Limb Darkening\label{sec_LD}}

Using the coefficients found in \citet{2000MNRAS.318..387D},
a lookup table was generated to convert our initial uniform disk measurements
of angular size into limb darkened disk measures appropriate for
consideration of parameters such as effective temperature and
linear radius.  \citet{2000MNRAS.318..387D} provide a grid of multiplicative
conversion factors, $\theta_{\rm LD} / \theta_{\rm UD}$, derived from Kurucz model stellar atmospheres.
To utilize the data found in this work, the data relevant to PTI's
near-infrared bandpasses was extracted to synthesize the aggregate
effect across the broadband filters of PTI.  From the uniform disk
diameter of a star, combined with its bolometric flux, a
rough estimate of its temperature was established.
Surface gravity, while provided in \citet{2000MNRAS.318..387D}, is not a
significant factor in establishing the conversion factor in
the $\log(g) = 0$ to $1$ range relevant for the supergiants being
investigated in this work.  The temperature-dependent
conversion factors are presented in Table \ref{table_LDUD}.
Based on that
temperature estimate, a $\theta_{\rm LD} / \theta_{\rm UD}$ cofficient
was established, and an estimate of $\theta_{\rm LD}$ was derived from
$\theta_{\rm UD}$.

Although this approach could be considered circular -
the $\theta_{\rm LD} / \theta_{\rm UD}$ coefficients are selected from
an initial estimate of $T$ from  $\theta_{\rm UD}$, and then are used to
revise the estimate of $T$ - the coefficients are sufficiently small
in this application that a more proper, recursive approach is not necessary.  In
general, the application of the $\theta_{\rm LD} / \theta_{\rm UD}$ coefficients
revise the stellar effective temperatures downward by $\Delta T = -56$K, a
shift that is sufficiently fine with respect to the $\theta_{\rm LD} / \theta_{\rm UD}$
grid that no further consideration is necessary.

\input{tab_UD2LD.tex}

\subsection{Dereddened Stellar Colors}

Given the value for $A_{\rm V}$ derived during the SED fit in \S \ref{sec_sedFitting}, we
may assume a standard $R_{\rm V}=3.1$ wavelength progression of reddening to establish values
for K-band reddening ($A_{\rm K} = 0.11 A_{\rm V}$) and reddening at 12 $\mu$m ($A_{\rm [12]} = 0.028 A_{\rm V}$)
\citep{1989ApJ...345..245C}.  From these reddening parameters, `true' values for $V_0$, $K_0$, and $[12]_0$ were established
for the stars in this investigation.

\section{Observed Stellar Parameters}
\subsection{Effective Temperature versus V-K Color}

Stellar effective temperature ($T_{\rm EFF}$) is defined in terms of the star's luminosity and radius by the Stefan-Boltzmann equation, $L = 4\pi \sigma R^2T_{\rm EFF}^4$ \citep{Stefan1879, Boltzmann1884}. Rewriting this equation in terms of angular diameter $\theta_{\rm LD}$ and bolometric flux $F_{\rm BOL}$, $T_{\rm EFF}$ can be expressed as $T_{\rm EFF} \propto (F_{\rm BOL}/\theta^2_{\rm LD})^{1/4}$. The limb darkened angular size is utilized here as a reasonable proxy for the Rosseland angular diameter, which corresponds to the surface where the Rosseland mean optical depth equals unity, as advocated by \citet{1987A&A...186..200S} as the most appropriate surface for computing an effective temperature.

\begin{figure*}
\begin{center}
\includegraphics[scale=0.80,angle=0]{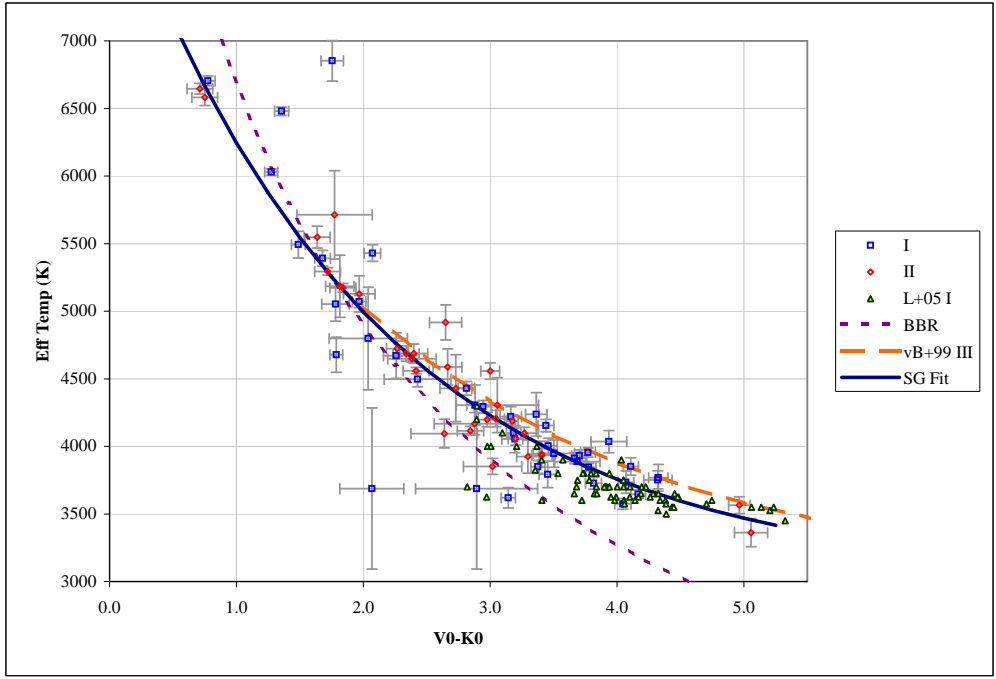}
\end{center}
\caption{\label{fig_teff_vs_V0K0} Effective temperature versus $(V-K)_0$ color ($V-K$ dereddened
for interstellar extinction) for luminosity class I and II stars.  Temperature-color values from \citet{2005ApJ...628..973L}
are included for comparison, as is the relationship for luminosity class III stars from \citet{1999AJ....117..521V} (dashed line) and for blackbody radiators (dotted line).  The fit for the stars given in this paper is shown as well (solid line).}
\end{figure*}

\begin{figure*}
\begin{center}
\includegraphics[scale=0.80,angle=0]{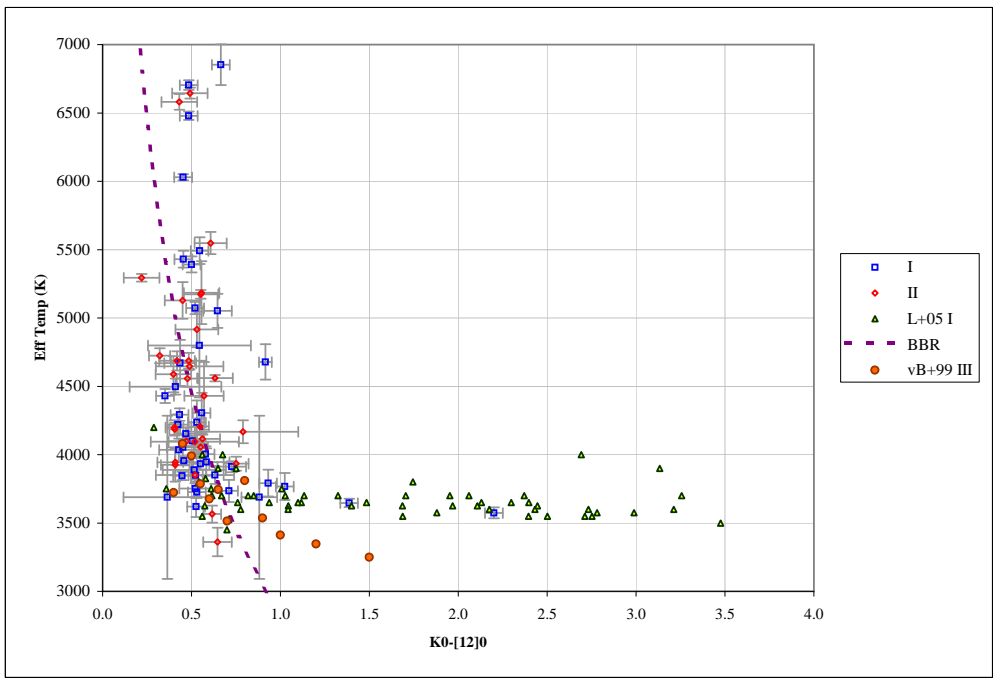}
\end{center}
\caption{\label{fig_teff_vs_K0120} Effective temperature versus $K_0-[12]_0$ color ($K-[12]$ dereddened
for interstellar extinction) for luminosity class I and II stars.  Temperature-color values from \citet{2005ApJ...628..973L}
are included for comparison, as is the relationship for luminosity class III stars from \citet{1999AJ....117..521V} and for blackbody radiators (dotted line).}
\end{figure*}

The resulting values are plotted as a function of $(V - K)_0$ and $(K - [12])_0$ colors in Figures \ref{fig_teff_vs_V0K0} and \ref{fig_teff_vs_K0120}, respectively.  For the first color, a fit of the data for both luminosity classes produces the relationship:
\begin{equation}
T_{\rm EFF} = 3037 (\pm 89) + 5264 (\pm 80) \times 10^{-0.2158 (\pm 0.0014) \times (V - K)_0}
\end{equation}
with a reduced chi-squared value of $\chi^2_\nu = 2.90$.
This empirical calibration has a median deviation of $\Delta T = 70$K in the range of
$0.7 < V-K < 5.1$ for LC I stars; for LC II, the median deviation is $\Delta T =
120$K from $0.4 < V-K < 4.3$.

Of particular interest is comparison of these results with those for giant stars found in \citet{1999AJ....117..521V}.  At first inspection, it appears the supergiants are, for a given temperature, slightly bluer than the giant stars.  However, this appears within the statistical uncertainty; it is also worth noting that (due to the modeling constraints of the time) the approach to reddening in the giant star paper is markedly less sophisticated than the one employed here.  It is our expectation that improved values for $A_{\rm V}$ for the giant stars will have the potential to shift the relationship derived in that paper to the left, overlapping with the one established here.

For comparison, we may examine the data found in \citet{2005ApJ...628..973L}, derived spectroscopically for supergiants.  The
color-temperature points found in that investigation are also plotted in Figure \ref{fig_teff_vs_V0K0}, and upon
inspection of the solid line fit to the PTI data, appear to be cooler.  However, the average value of the disagreement is $\Delta T=60\pm85$K overall, which is not statistically significant; for those stars in the range of $3.0 < (V - K)_0 < 4.5$, that disagreement is $\Delta T = 75 \pm 70$K, which is a bit more meaningful, but only slightly so.

For temperature as a function of $(K - [12])_0$, Figure \ref{fig_teff_vs_K0120} exhibits a tendency seen previously in \citet{1999AJ....117..521V} for giant stars: as $T_{\rm EFF}$ approaches 3600-3700K, the $(K - [12])_0$ color increases rapidly away from the nominal blackbody relationship, indicative of dusty mass loss \citep{1990AJ.....99.1569B,1996A&A...314..896L}.  The hotter objects in excess of $\sim$4750K also appear to track rightwards of the blackbody line, also suggesting mass loss from these objects.

\subsection{Effective Temperature versus Spectral Type\label{sec_teff_sptype}}

\begin{figure*}
\begin{center}
\includegraphics[scale=0.80,angle=0]{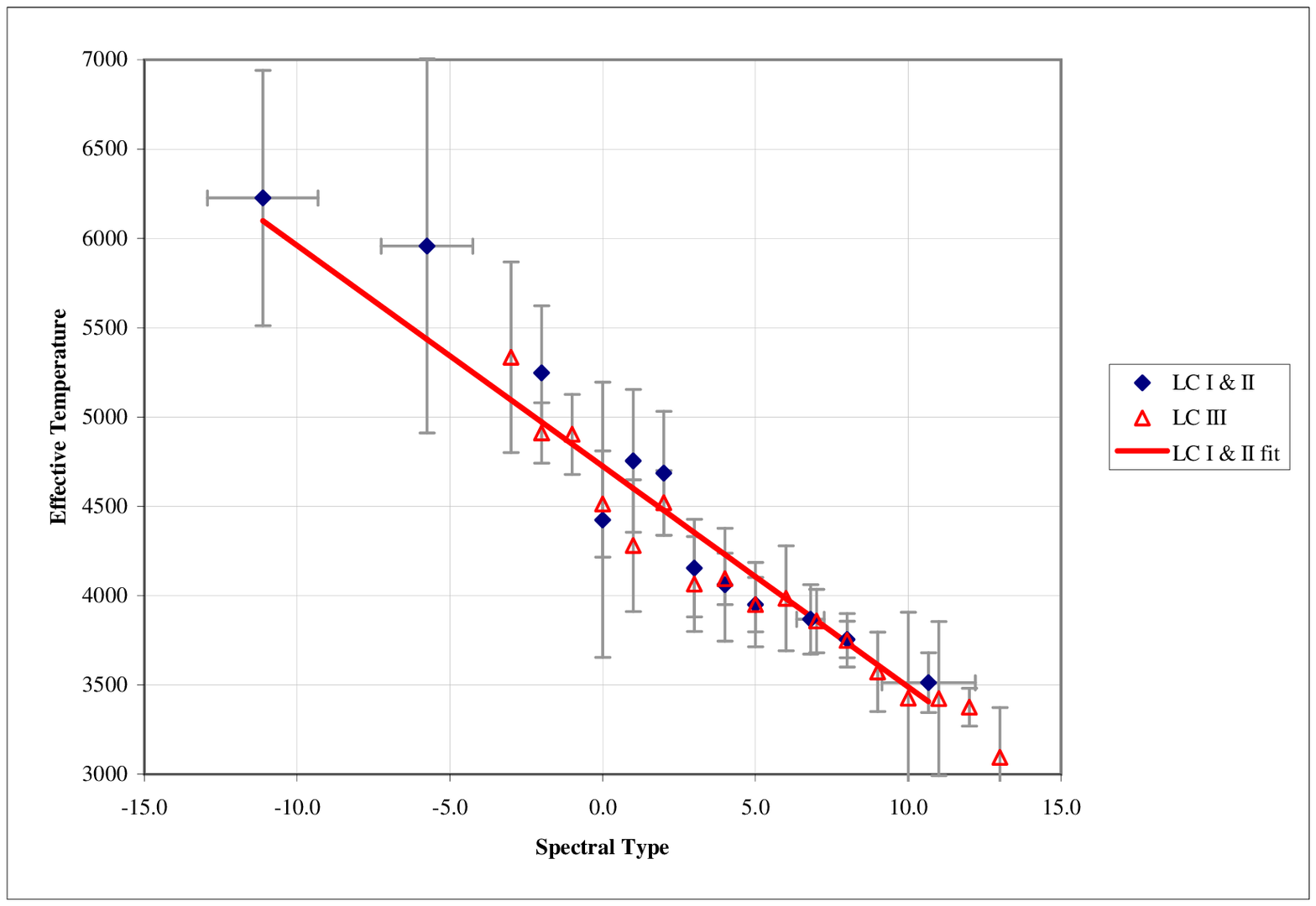}
\end{center}
\caption{\label{fig_T_vs_sptype} Binned effective temperature versus spectral type, with $ST=-2, \ldots 0, \ldots 5, 6, \ldots 14$ corresponding to the spectral types ${\rm G8}, \ldots {\rm K0}, \ldots {\rm K5, M0}, \ldots {\rm M8}$ (as discussed in \S \ref{sec_teff_sptype}) for luminosity class I \& II stars, along with class III data from \citet{1999AJ....117..521V} and a fit to the class I \& II objects.  No evidence is seen for a temperature offset between LC I\&II and LC III.}
\end{figure*}

\citet{1998AJ....116..981D} noted that there appeared to be observational evidence that suggested that, for a given spectral type, supergiants were systematically cooler than their giant star counterparts.  To explore this possibility, the data were binned by spectral type as seen in Table \ref{table_teff_sptype}.  A fit to these binned data is
\begin{equation}
T_{\rm EFF} = -123 (\pm25) \times ST + 4724 (\pm175)
\end{equation}
with a reduced $\chi^2_\nu$ of 0.34, where the spectral type $ST=-2, \ldots 0, \ldots 5, 6, \ldots 14$ corresponding to ${\rm G8}, \ldots {\rm K0}, \ldots {\rm K5, M0}, \ldots {\rm M8}$ as in \citet{1998AJ....116..981D}.  This fit, the binned data, and the corresponding giant star data from \citet{1998AJ....116..981D} are seen in Figure \ref{fig_T_vs_sptype}.  
The small value of $\chi^2_\nu$ is indicative of the large errors for each bin found in Table \ref{table_teff_sptype}; this is possibly traceable to a degree of natural spread found in $T_{\rm EFF}$ within each spectral type.
Comparing this fit to the fit in \citet{1998AJ....116..981D}, $T_{\rm EFF} = -109 \times ST + 4570$ K, we find that it is statistically identical: there is no evidence for temperature deviation by spectral type between luminosity classes.

By spectral bin, the absolute average deviation from the fit is $\Delta T \approx 140$ K, which is substantially smaller than the standard deviations by spectral type bin - on average, $\Delta T \approx 400$ K - seen in Table \ref{table_teff_sptype} (and hence the small value for $\chi^2_\nu$).  This large scatter suggests it is impractical to expect any degree of precision in predicting the effective temperature of bright giants and supergiants from their spectral types, and that employing the $V-K$ index is substantively more useful.

Also in Table \ref{table_teff_sptype}, for reference, are data points from \citet{2005ApJ...628..973L} moderate resolution optical spectroscopy.  Agreement is good with the exception of the K-type stars. The ambiguity in these particular data points is reflected in the error bars, which are substantially larger than those for the cooler stars published in \citet{2005ApJ...628..973L}. We have confidence in our numbers because the temperature trend continues to much earlier spectral types.

\input{tab_teff_sptype.tex}

\subsection{Linear Radius\label{sec_radius}}

\begin{figure*}
\begin{center}
\includegraphics[scale=0.80,angle=0]{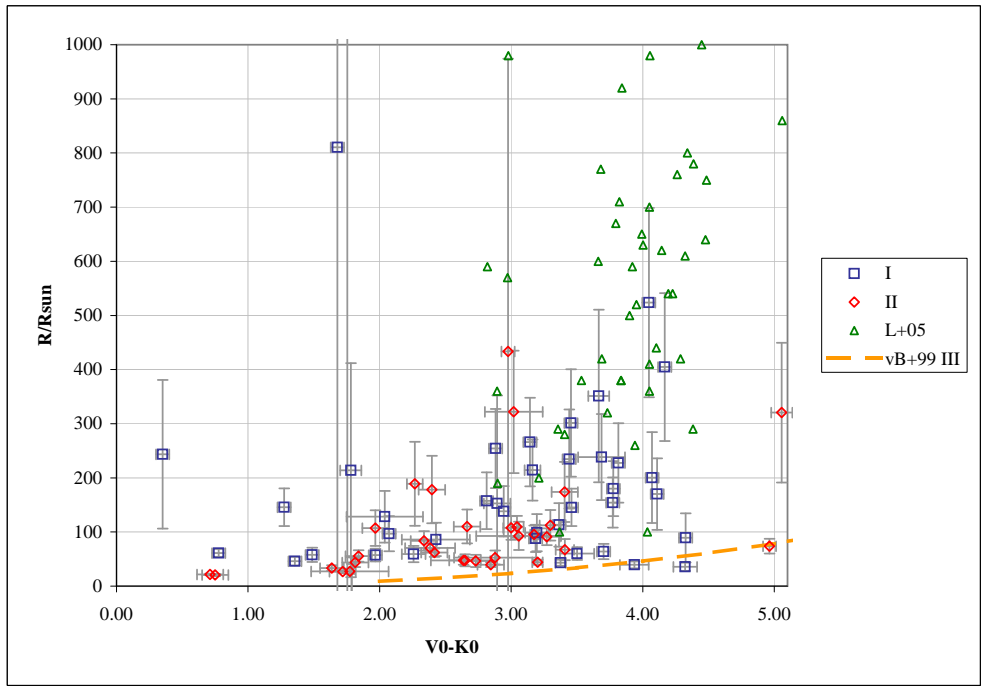}
\end{center}
\caption{\label{fig_R_vs_V0K0} Linear radius versus $(V-K)_0$ color ($V-K$ dereddened
for interstellar extinction) for luminosity class I and II stars.  Radius-color values from \citet{2005ApJ...628..973L}
are included for comparison, as is the relationship for luminosity class III stars from \citet{1999AJ....117..521V} (dashed line).}
\end{figure*}

Stellar radius is proportional to angular size and distance, $R \propto \theta_{\rm LD} d$, and is plotted in Figure \ref{fig_R_vs_V0K0} versus $(V-K)_0$, and given by $(V-K)_0$ bin in Table \ref{table_R}.
Unfortunately, due to the large errors in the distance measurements, there is considerable scatter in these data.  In fact both the weighted and unweighted radii are presented in Table \ref{table_R}, to give a sense of any possible Lutz-Kelker bias\footnote{A systematic error that causes observed parallaxes to be too large \citep{1973PASP...85..573L}. See the discussion in \citet{1998MNRAS.294L..41O} for more details.} that may be influencing the results.
There are a few class I objects in the range $3.5 < (V-K)_0 < 4.5$ that do appear unusually small (HD 30178, HD 31220, HD 44033, HD 200576, HD 207991, HD 236915); so much so, and in agreement with fit line for the giants, that it is worth considering whether or not these objects have been misclassified\footnote{This appears plausible.  With the exception of HD 207991, the spectral typings for these objects are traceable to older, objective-prism typings, which might have led to erroneous luminosity classifications.}.  These objects have been excluded from the bins in Table \ref{table_R}.

However, some features are readily apparent in a qualitative sense:  First, the trend is for the luminosity class I and II objects to be, on average, larger than the giants of \citet{1999AJ....117..521V}.  Second, the luminosity class I objects appear on average to be larger than the class II objects.  Both of these qualitative observations are expected from luminosity properties \citep{1987A&A...177..217D} and temperatures of these stars.

Quantitatively, comparing the general trends by $(V-K)_0$ bin in Table \ref{table_R}, and in \citet{1999AJ....117..521V}, we can infer a general rule: for a given $(V-K)_0$ color, a luminosity class I object will be roughly $2 \times$ larger than a class II counterpart, just as a class II object will be twice as large as a giant of similar color.  This rule appears to be supported by the data over the range of $2.0 < (V-K)_0 < 3.5$.  A simple examination of the implication of this rule for the over luminosities of class I and II objects, compared to the determinations of \citet{1987A&A...177..217D}, show it to be not in perfect agreement: the expectation is that, for a constant temperature by spectral type, the radii should scale by a factor of roughly $4 \times$, not $2 \times$, between the luminosity classes.  Again, this disagreement is possibly due to the inaccuracies of the distance measurements for these objects, or possibly outright bias in those distance determinations.  As mentioned in \S \ref{sec_radius}, if there were significant Lutz-Kelker bias for a large portion of the sample, the actual distances would on average be greater than currently indicated.

Finally, the region of phase space occupied by the supergiants from \citet{2005ApJ...628..973L} appear to be redder and linearly larger than those osberved with PTI.  This is consistent with these objects undergoing a higher rate of mass loss, as previously discussed with regards to Figure \ref{fig_teff_vs_K0120}.

\input{tab_R.tex}

\section{Conclusions}

The 74 bright giants and supergiants presented herein represent a large, homogenous data set exploring the parameters of effective temperature and linear radius for these, the largest of stars.  Empirical calibrations for $T_{\rm EFF}$  versus $(V-K)_0$ and $T_{\rm EFF}$ versus spectral type for these stars are established, and is found to be in agreement with similar relationships explored for giant stars.  Less meaningful statements can be made about their linear radii, due to the poor quality of distance information for these objects; however, the progression in linear size from giant stars to bright giants to supergiants is consistent with qualitative expectations.  Improved distance estimates from upcoming missions such as the {\it Space Interferometry Mission} (SIM) \citep{2008PASP..120...38U} or {\it Gaia} \citep{2007ecf..book..205G} have the potential to greatly leverage this data for additional insight into the fundamental physical nature of these stars.

\input{tab_starResults_LC1.tex}

\input{tab_starResults_LC2.tex}

\section*{Acknowledgments}

We would like to acknowledge the helpful comments of a referee, which helped to improve the quality of this manuscript.
Science operations with PTI are conducted through the efforts of the PTI Collaboration (http://pti.jpl.nasa.gov/ptimembers.html), and we acknowledge the invaluable contributions of our PTI colleagues.
We particularly thank Kevin Rykoski for his professional (and frequently heroic) operation of PTI.
The Palomar Testbed Interferometer is operated by the Michelson Science Center on and the PTI collaboration and was constructed with funds from the Jet Propoulsion Laboratory, Caltech, as provided by the National Aeronautics and Space Administration. This research has made use of services produced by the Michelson Science Center at the California Institute of Technology, the SIMBAD database, operated at CDS, Strasbourg, France, and data products from the Two Micron All Sky Survey, which is a joint project of the University of Massachusetts and the Infrared Processing and Analysis Center/California Institute of Technology, funded by the National Aeronautics and Space Administration and the National Science Foundation.  Portions of this work were performed at the Jet Propulsion Laboratory, California Institute of Technology under contract with the National Aeronautics and Space Administration.

\label{lastpage}

\end{document}

%% file: journal_defs.tex
\def\aj{AJ}				
\def\araa{ARA\&A}			
\def\apj{ApJ}				
\def\apjl{ApJ}				
\def\apjs{ApJS}				 
\def\aap{A\&A}				
\def\aaps{A\&AS}			
\def\bams{\ref@jnl{Bull.~Am.~Meteorol.~Soc.}} 
\def\bssa{\ref@jnl{Bull.~Seismol.~Soc.~Am.}} 
\def\dsr{\ref@jnl{Deep~Sea~Res.}}	
\def\eos{\ref@jnl{Eos~Trans.~AGU}}	
\def\epsl{\ref@jnl{Earth~Planet.~Sci.~Lett.}}	
\def\gca{\ref@jnl{Geochim.~Cosmochim.~Acta}}	
\def\gjras{\ref@jnl{Geophys.~J.~R.~Astron.~Soc.}} 
\def\grl{\ref@jnl{Geophys.~Res.~Lett.}}	
\def\gsab{\ref@jnl{Geol.~Soc.~Am.~Bull.}}	
\def\jatp{\ref@jnl{J.~Atmos.~Terr.~Phys.}}	
\def\jgr{\ref@jnl{J.~Geophys.~Res.}}	
\def\jpo{\ref@jnl{J.~Phys.~Oceanogr.}}	
\def\mnras{MNRAS}			
\def\mwr{\ref@jnl{Mon.~Weather~Rev.}}	

\def\pepi{\ref@jnl{Phys.~Earth Planet.~Inter.}}	
\def\pra{\ref@jnl{Phys.~Rev.~A}}		
\def\prb{\ref@jnl{Phys.~Rev.~B}}		
\def\prc{\ref@jnl{Phys.~Rev.~C}}		
\def\prd{\ref@jnl{Phys.~Rev.~D}}		
\def\prl{\ref@jnl{Phys.~Rev.~Lett}}	
\def\pasp{PASP}				
\def\qjrms{\ref@jnl{Q.~J.~R.~Meteorol.~Soc.}}	
\def\rg{\ref@jnl{Rev.~Geophys.}}	
\def\rs{\ref@jnl{Radio~Sci.}}		
\def\usgsof{\ref@jnl{U.S.~Geol.~Surv. Open File~Rep.}}	
\def\usgspp{\ref@jnl{U.S.~Geol.~Surv.~Prof.~Pap.}}	

%% file: tab_UD2LD.tex
\begin{table*}
 \centering
 \begin{minipage}{140mm}
   \caption{K-band uniform disk-to-limb darkened disk multiplication coefficients
extracted from \citet{2000MNRAS.318..387D}.\label{table_LDUD}}
  \begin{tabular}{@{}cccc@{}}
  \hline
{$T$} &
{$\log(g)=0.0$} &
{$\log(g)=1.0$} &
{Average} \\
{(K)} &
{$\theta_{\rm LD} / \theta_{\rm UD}$} &
{$\theta_{\rm LD} / \theta_{\rm UD}$} &
{$\theta_{\rm LD} / \theta_{\rm UD}$}\\
 \hline
3500 & 1.0307 & 1.0303 & 1.0305 \\
3750 & 1.0295 & 1.0293 & 1.0294 \\
4000 & 1.0281 & 1.0279 & 1.0280 \\
4250 & 1.0265 & 1.0264 & 1.0264 \\
4500 & 1.0248 & 1.0247 & 1.0248 \\
4750 & 1.0231 & 1.0231 & 1.0231 \\
5000 & 1.0214 & 1.0216 & 1.0215 \\
5250 & 1.0197 & 1.0199 & 1.0198 \\
5500 & 1.0206 & 1.0181 & 1.0193 \\
5750 & 1.0197 & 1.0165 & 1.0181 \\
6000 & 1.0193 & 1.0184 & 1.0189 \\
6250 & n/a & 1.0176 & 1.0176 \\
6500 & n/a & 1.0170 & 1.0170 \\
6750 & n/a & 1.0166 & 1.0166 \\
\hline
\end{tabular}
\end{minipage}
\end{table*}

%% file: tab_teff_sptype.tex
\begin{table*}
 \centering
 \begin{minipage}{140mm}
   \caption{Effective temperature by spectral type bins.  As in \citet{1998AJ....116..981D}, spectral type $-2, \ldots 0, \ldots 5, 6, \ldots 14$ corresponds to ${\rm G8}, \ldots {\rm K0}, \ldots {\rm K5, M0}, \ldots {\rm M8}$.\label{table_teff_sptype}}
  \begin{tabular}{@{}ccccccc@{}}
  \hline
{Lower} &
{Upper} &
{$N$} &
{Average} &
{Average} &
Binned&
\citet{2005ApJ...628..973L}\\
{Sptype} &
{Sptype} &
  &
{SpType Num}  &
{Sptype} &
{$T_{\rm EFF}$} (K)&
{$T_{\rm EFF}$}(K)\\
 \hline
-15.0 & -9.5 & 8 & $-11.1 \pm 1.8 $ &F9& $6226 \pm 715$ & \\
-9.5 & -4.5 & 4 & $-5.8 \pm 1.5 $ &G4& $5958 \pm 1047$ & \\
-4.5 & -1.5 & 4 & $-2.0 \pm 0.0 $ &G8& $5247 \pm 375$ & \\
-1.5 & 0.5 & 4 & $0.0 \pm 0.0  $ &K0& $4424 \pm 770$    & \\
0.5 & 1.5 & 4 & $1.0 \pm 0.0  $ &K1& $4755 \pm 399$   & $4150 \pm 210$\\
1.5 & 2.5 & 6 & $2.0 \pm 0.0  $ &K2& $4685 \pm 346$    & $4075 \pm 100$\\
2.5 & 3.5 & 11 & $3.0 \pm 0.0  $ &K3& $4154 \pm 273$    & $4000$ \\
3.5 & 4.5 & 4 & $4.0 \pm 0.0 $ &K4 & $4062 \pm 316$   & \\
4.5 & 5.5 & 13 & $5.0 \pm 0.0 $ &K5 & $3950 \pm 235$   & 3900\\
5.5 & 7.5 & 5 & $6.8 \pm 0.4 $ &M0.8& $3867 \pm 194$    & $3740 \pm 45$\\
7.5 & 8.5 & 7 & $8.0 \pm 0.0 $ &M2 &  $3755 \pm 103$    & $3660 \pm 30$\\
8.5 & 12.5 & 3 & $10.7 \pm 1.5 $ &M4.7& $3513 \pm 168$    & $3525\pm 35 $ \\
\hline
\end{tabular}
\end{minipage}
\end{table*}

%% file: tab_R.tex
\begin{table*}
 \centering
 \begin{minipage}{140mm}
   \caption{Linear radii as a function of $V_0 - K_0$ color for luminosity class I and II stars.  Sample size $N$
is given for each luminosity class, along with weighted and (unweighted) averages.  The six stars discussed in \S \ref{sec_radius} are exempted from this table.\label{table_R}}
  \begin{tabular}{@{}ccccc@{}}
  \hline
{$V_0-K_0$} &
{$N$ (I)} &
{$R$ (I)} &
{$N$ (II)} &
{$R$ (II)}\\
 \hline
$\{0,1\}$ & 2 & 62 (152) & 2 & 21 (21) \\
$\{1,2\}$ & 9 & 54 (15) & 6 & 30 (49) \\
$\{2,3\}$ & 8 & 85 (134) & 13 & 52 (113) \\
$\{3,4\}$ & 13 & 141 (201) & 9 & 58 (123) \\
$\{4,5\}$ & 4 & 232 (325) & 1 & 74 (74) \\
$\{5,6\}$ & 0 &   & 1 & 321 (321) \\
\hline
\end{tabular}
\end{minipage}
\end{table*}

%% file: tab_starResults_LC1.tex
\begin{table*}
 \centering
 \begin{minipage}{200mm}
   \caption{Unreddened colors, effective temperatures and radii for luminosity class I stars.\label{table_LC1results}.}
  \begin{tabular}{@{}ccccccc@{}}
  \hline
{Star ID} &
{Spectral} &
{$V_0-K_0$} &
{$K_0-[12]_0$} &
{$T_{\rm EFF}$} &
{$d$} &
{$R$} \\
 &
{Type} &
{(mag)} &
{(mag)} &
{(K)} &
{(pc)} &
{$(R_\odot)$}\\
 \hline
HD 4817 & K5Ib & $3.81 \pm 0.06$ & $0.53 \pm 0.05$ & $3727 \pm 44$ & $762 \pm 245$ & $227 \pm 73$ \\
HD 9366 & K3Ib & $2.88 \pm 0.07$ & $0.56 \pm 0.05$ & $4305 \pm 148$ & $1472 \pm 412$ & $254 \pm 73$ \\
HD 9900 & K1Ia & $2.26 \pm 0.10$ & $0.44 \pm 0.09$ & $4671 \pm 169$ & $350 \pm 86$ & $59 \pm 15$ \\
HD 11092 & K5Ia0-a... & $3.16 \pm 0.08$ & $0.42 \pm 0.06$ & $4220 \pm 74$ & $732 \pm 193$ & $214 \pm 56$ \\
HD 12399 & G5Ia & $1.79 \pm 0.05$ & $0.92 \pm 0.04$ & $4678 \pm 129$ & $-1389 \pm 1755$ & $-155 \pm 196$ \\
HD 13658 & M1Iab & $3.45 \pm 0.07$ & $0.93 \pm 0.05$ & $3792 \pm 98$ & $1852 \pm 603$ & $301 \pm 99$ \\
HD 13686 & K3Ib & $3.20 \pm 0.10$ & $0.45 \pm 0.09$ & $4054 \pm 101$ & $597 \pm 206$ & $99 \pm 34$ \\
HD 14404 & M2Ib & $4.17 \pm 0.06$ & $1.39 \pm 0.05$ & $3647 \pm 30$ & $1770 \pm 597$ & $405 \pm 137$ \\
HD 17958 & K3Ibvar & $2.81 \pm 0.07$ & $0.35 \pm 0.05$ & $4429 \pm 52$ & $533 \pm 178$ & $158 \pm 53$ \\
HD 18391 & G0Ia & $1.78 \pm 0.11$ & $0.65 \pm 0.08$ & $5052 \pm 125$ & $1099 \pm 1014$ & $214 \pm 198$ \\
HD 20902 & F5Ib & $0.78 \pm 0.06$ & $0.48 \pm 0.05$ & $6704 \pm 36$ & $181 \pm 22$ & $61 \pm 7$ \\
HD 21465 & K5Iab: & $3.36 \pm 0.08$ & $0.53 \pm 0.05$ & $4236 \pm 160$ & $574 \pm 200$ & $113 \pm 40$ \\
HD 30178 & M2Ib & $4.32 \pm 0.11$ & $0.52 \pm 0.09$ & $3752 \pm 66$ & $185 \pm 40$ & $36 \pm 8$ \\
HD 31220 & M0Ib & $3.70 \pm 0.06$ & $0.55 \pm 0.05$ & $3933 \pm 42$ & $283 \pm 62$ & $64 \pm 14$ \\
HD 31910 & G0Ib & $1.49 \pm 0.05$ & $0.55 \pm 0.05$ & $5493 \pm 98$ & $306 \pm 69$ & $58 \pm 13$ \\
HD 34255 & K4Iab: & $3.46 \pm 0.07$ & $0.58 \pm 0.05$ & $4004 \pm 56$ & $523 \pm 124$ & $145 \pm 35$ \\
HD 35601 & M1Ib & $3.67 \pm 0.09$ & $0.73 \pm 0.08$ & $3912 \pm 41$ & $1158 \pm 525$ & $351 \pm 159$ \\
HIP 35915 & M6I-II & $2.89 \pm 0.48$ & $0.88 \pm 0.10$ & $3688 \pm 597$ & $730 \pm 954$ & $153 \pm 199$ \\
HD 37387 & K1Ib & $2.04 \pm 0.31$ & $0.54 \pm 0.29$ & $4798 \pm 379$ & $856 \pm 298$ & $128 \pm 48$ \\
HD 38247 & G8Iab & $2.07 \pm 0.25$ & $0.36 \pm 0.25$ & $3688 \pm 597$ &  &  \\
HD 44033 & K3Ib & $3.38 \pm 0.06$ & $0.52 \pm 0.04$ & $3849 \pm 48$ & $169 \pm 22$ & $43 \pm 6$ \\
HD 52005 & K3Ib & $3.14 \pm 0.06$ & $0.53 \pm 0.05$ & $3619 \pm 76$ & $875 \pm 267$ & $266 \pm 82$ \\
HD 183589 & K5Ib & $3.77 \pm 0.06$ & $0.46 \pm 0.05$ & $3955 \pm 34$ & $527 \pm 159$ & $154 \pm 46$ \\
HD 186021 & K0Iab: & $2.43 \pm 0.26$ & $0.41 \pm 0.26$ & $4498 \pm 58$ & $623 \pm 222$ & $86 \pm 31$ \\
HD 187238 & K3Ia0-a... & $2.94 \pm 0.07$ & $0.43 \pm 0.05$ & $4293 \pm 46$ & $686 \pm 231$ & $138 \pm 47$ \\
HD 193469 & K5Ib & $3.78 \pm 0.06$ & $0.45 \pm 0.05$ & $3845 \pm 30$ & $714 \pm 201$ & $180 \pm 51$ \\
HD 194093 & F8Ib & $1.27 \pm 0.05$ & $0.45 \pm 0.05$ & $6029 \pm 23$ & $467 \pm 111$ & $146 \pm 35$ \\
HD 196725 & K3Ib & $3.19 \pm 0.06$ & $0.50 \pm 0.05$ & $4100 \pm 36$ & $480 \pm 137$ & $88 \pm 25$ \\
HD 197345 & A2Ia & $0.35 \pm 0.05$ & $-2.47 \pm 0.05$ & $8211 \pm 28$ & $990 \pm 559$ & $243 \pm 137$ \\
HD 200576 & K5Ib & $3.50 \pm 0.16$ & $0.58 \pm 0.13$ & $3947 \pm 101$ & $368 \pm 70$ & $60 \pm 12$ \\
HD 205349 & K1Ibvar & $2.07 \pm 0.07$ & $0.45 \pm 0.05$ & $5430 \pm 62$ & $585 \pm 196$ & $97 \pm 33$ \\
HD 206859 & G5Ib & $1.97 \pm 0.05$ & $0.52 \pm 0.05$ & $5072 \pm 45$ & $276 \pm 56$ & $57 \pm 12$ \\
HD 207119 & K5Ib & $3.44 \pm 0.06$ & $0.47 \pm 0.05$ & $4154 \pm 46$ & $968 \pm 378$ & $235 \pm 92$ \\
HD 207991 & K5Ib & $3.94 \pm 0.14$ & $0.43 \pm 0.11$ & $4035 \pm 81$ & $316 \pm 52$ & $39 \pm 6$ \\
HD 213306 & G2Ibvar & $1.36 \pm 0.05$ & $0.48 \pm 0.05$ & $6479 \pm 31$ & $301 \pm 53$ & $46 \pm 8$ \\
HD 216946 & K5Ibvar & $3.69 \pm 0.18$ & $0.52 \pm 0.18$ & $3888 \pm 19$ & $575 \pm 192$ & $239 \pm 80$ \\
HD 217476 & G0Ia & $1.76 \pm 0.09$ & $0.66 \pm 0.05$ & $6853 \pm 150$ & $-6667 \pm 25333$ & $-878 \pm 3337$ \\
HD 224014 & F8Iavar & $1.68 \pm 0.06$ & $0.50 \pm 0.05$ & $5392 \pm 59$ & $3571 \pm 7398$ & $811 \pm 1679$ \\
HD 236697 & M2Ib & $4.11 \pm 0.06$ & $0.63 \pm 0.05$ & $3851 \pm 64$ & $1199 \pm 465$ & $170 \pm 66$ \\
HD 236915 & M2Iab & $4.33 \pm 0.08$ & $1.02 \pm 0.05$ & $3768 \pm 99$ & $422 \pm 214$ & $89 \pm 45$ \\
HD 236947 & M2Ia0-a... & $4.07 \pm 0.06$ & $0.71 \pm 0.05$ & $3735 \pm 81$ & $1167 \pm 486$ & $200 \pm 84$ \\
HD 236979 & M1Iab:var & $4.05 \pm 0.06$ & $2.20 \pm 0.05$ & $3574 \pm 40$ & $1780 \pm 594$ & $524 \pm 175$ \\
\hline
\end{tabular}
\end{minipage}
\end{table*}

%% file: tab_starResults_LC2.tex
\begin{table*}
 \centering
 \begin{minipage}{200mm}
   \caption{Unreddened colors, effective temperatures and radii for luminosity class II stars.\label{table_LC2results}.}
  \begin{tabular}{@{}ccccccc@{}}
  \hline
{Star ID} &
{Spectral} &
{$V_0-K_0$} &
{$K_0-[12]_0$} &
{$T_{\rm EFF}$} &
{$d$} &
{$R$} \\
 &
{Type} &
{(mag)} &
{(mag)} &
{(K)} &
{(pc)} &
{$(R_\odot)$}\\
 \hline
HD 1613 & M2II: & $3.02 \pm 0.23$ & $0.52 \pm 0.22$ & $3852 \pm 59$ & $1185 \pm 416$ & $322 \pm 113$ \\
HD 3147 & K2Ib-II & $2.40 \pm 0.11$ & $0.42 \pm 0.10$ & $4687 \pm 70$ & $845 \pm 295$ & $178 \pm 62$ \\
HD 3489 & K3Ib-II & $3.41 \pm 0.11$ & $0.41 \pm 0.10$ & $3943 \pm 95$ & $1177 \pm 372$ & $174 \pm 56$ \\
HD 8701 & K2II:p & $1.97 \pm 0.12$ & $0.45 \pm 0.10$ & $5128 \pm 134$ & $684 \pm 207$ & $107 \pm 33$ \\
HD 13725 & K4II & $2.66 \pm 0.11$ & $0.40 \pm 0.10$ & $4588 \pm 133$ & $663 \pm 186$ & $110 \pm 31$ \\
HD 20123 & G5II & $1.81 \pm 0.11$ & $0.56 \pm 0.10$ & $5185 \pm 230$ & $256 \pm 44$ & $44 \pm 8$ \\
HD 22135 & K5II & $2.73 \pm 0.13$ & $0.57 \pm 0.11$ & $4430 \pm 247$ & $290 \pm 65$ & $47 \pm 11$ \\
HD 28487 & M3II & $4.96 \pm 0.08$ & $0.62 \pm 0.05$ & $3565 \pm 62$ & $244 \pm 45$ & $74 \pm 14$ \\
HD 29094 & G8II & $1.84 \pm 0.10$ & $0.55 \pm 0.10$ & $5173 \pm 32$ & $197 \pm 38$ & $56 \pm 11$ \\
HD 30504 & K4II & $3.20 \pm 0.06$ & $0.55 \pm 0.04$ & $4056 \pm 47$ & $150 \pm 17$ & $44 \pm 5$ \\
HD 31767 & K2IIvar & $2.34 \pm 0.10$ & $0.48 \pm 0.10$ & $4687 \pm 57$ & $300 \pm 64$ & $84 \pm 18$ \\
HD 39225 & M2II: & $3.41 \pm 0.08$ & $0.75 \pm 0.07$ & $3934 \pm 51$ & $262 \pm 76$ & $67 \pm 19$ \\
HD 77912 & G8Ib-II & $1.64 \pm 0.10$ & $0.61 \pm 0.09$ & $5548 \pm 81$ & $208 \pm 33$ & $33 \pm 5$ \\
HD 84441 & G0II & $0.71 \pm 0.10$ & $0.49 \pm 0.10$ & $6645 \pm 40$ & $77 \pm 5$ & $21 \pm 1$ \\
HD 159968 & M1II & $3.27 \pm 0.11$ & $0.47 \pm 0.10$ & $4099 \pm 42$ & $290 \pm 49$ & $91 \pm 15$ \\
HD 163770 & K1IIvar & $2.38 \pm 0.19$ & $0.49 \pm 0.19$ & $4647 \pm 25$ & $213 \pm 25$ & $71 \pm 8$ \\
HD 180809 & K0II & $2.42 \pm 0.10$ & $0.63 \pm 0.10$ & $4560 \pm 23$ & $249 \pm 31$ & $62 \pm 8$ \\
HD 181475 & K5II & $2.98 \pm 0.07$ & $0.40 \pm 0.05$ & $4199 \pm 53$ & $1538 \pm 1917$ & $434 \pm 540$ \\
HD 185758 & G0II & $0.75 \pm 0.10$ & $0.43 \pm 0.10$ & $6581 \pm 59$ & $145 \pm 15$ & $21 \pm 2$ \\
HD 185958 & G8II & $1.72 \pm 0.10$ & $0.22 \pm 0.10$ & $5294 \pm 28$ & $143 \pm 14$ & $27 \pm 3$ \\
HD 191226 & K2II: & $2.27 \pm 0.09$ & $0.32 \pm 0.06$ & $4724 \pm 56$ & $1042 \pm 428$ & $189 \pm 78$ \\
HD 193092 & K5II & $3.17 \pm 0.05$ & $0.40 \pm 0.03$ & $4186 \pm 47$ & $333 \pm 56$ & $96 \pm 16$ \\
HD 193217 & K4II: & $3.30 \pm 0.09$ & $0.41 \pm 0.08$ & $3925 \pm 123$ & $590 \pm 144$ & $112 \pm 28$ \\
HD 196819 & K3II & $3.06 \pm 0.32$ &  & $4305 \pm 204$ & $733 \pm 193$ & $93 \pm 26$ \\
HD 199098 & K0II & $1.77 \pm 0.30$ &  & $5713 \pm 327$ & $249 \pm 38$ & $27 \pm 5$ \\
HD 213179 & K2II & $2.64 \pm 0.26$ & $0.52 \pm 0.25$ & $4095 \pm 107$ & $313 \pm 51$ & $47 \pm 8$ \\
HD 217673 & K2II & $2.65 \pm 0.13$ & $0.53 \pm 0.12$ & $4917 \pm 130$ & $406 \pm 90$ & $48 \pm 11$ \\
HD 218356 & K0IIp & $2.84 \pm 0.10$ & $0.56 \pm 0.10$ & $4115 \pm 33$ & $167 \pm 18$ & $40 \pm 4$ \\
HD 220369 & K3II & $3.00 \pm 0.07$ & $0.48 \pm 0.06$ & $4558 \pm 61$ & $468 \pm 95$ & $108 \pm 22$ \\
HD 223173 & K3II & $3.04 \pm 0.07$ & $0.55 \pm 0.05$ & $4206 \pm 60$ & $450 \pm 82$ & $109 \pm 20$ \\
HD 223332 & K5II & $2.88 \pm 0.32$ & $0.79 \pm 0.31$ & $4168 \pm 83$ & $438 \pm 106$ & $53 \pm 13$ \\
HD 334750 & M5II & $5.06 \pm 0.13$ & $0.65 \pm 0.08$ & $3362 \pm 104$ & $1324 \pm 532$ & $321 \pm 129$ \\\hline
\end{tabular}
\end{minipage}
\end{table*}

%% file: mn2_GvB_SG.bbl
\begin{thebibliography}{99}
\bibitem[Arellano Ferro et al.(1990)]{1990A&AS...83..225A} Arellano Ferro, A., Parrao, L., Schuster, W., Gonzalez-Bedolla, S., Peniche, R., \& Pena, J.~H.\ 1990, \aaps, 83, 225
\bibitem[Beichman et al.(1990)]{1990AJ.....99.1569B} Beichman, C.~A., Chester, T., Gillett, F.~C., Low, F.~J., Matthews, K., \& Neugebauer, G.\ 1990, \aj, 99, 1569
\bibitem[Bidelman(1957)]{1957PASP...69..147B} Bidelman, W.~P.\ 1957, \pasp, 69, 147
\bibitem[Boden et al.(1998)]{1998SPIE.3350..872B} Boden, A.~F., Colavita, M.~M., van Belle, G.~T., \& Shao, M.\ 1998, Proc. SPIE, 3350, 872
\bibitem[Boden et al.(1999)]{bod99} Boden, A.~F., et al. 1999, \apj, 515, 356
\bibitem[Boltzmann (1884)]{Boltzmann1884} Boltzmann, L., 1884, Annalen der Physik und Chemie, 22, 291
\bibitem[Boulon \& Fehrenbach(1958)]{1958JO.....42..149B} Boulon, J., \& Fehrenbach, C.\ 1958, Journal des Observateurs, 42, 149
\bibitem[Cardelli et al.(1989)]{1989ApJ...345..245C} Cardelli, J.~A., Clayton, G.~C., \& Mathis, J.~S.\ 1989, \apj, 345, 245
\bibitem[Colavita(1999)]{1999PASP..111..111C} Colavita, M.~M.\ 1999, \pasp, 111, 111
\bibitem[Colavita et al.(1999)]{col99}Colavita, M.M., et al., 1999, \apj, 510, 505
\bibitem[Cox(2000)]{cox00} Cox, A.~N.\ 2000, Allen's astrophysical quantities, 4th ed.~Publisher: New York: AIP Press; Springer, 2000.~Editedy by Arthur N.~Cox.~ ISBN: 0387987460,
\bibitem[Cutri et al.(2003)]{2003yCat.2246.....C} Cutri, R.~M., et al. 2003, VizieR Online Data Catalog, 2246,
\bibitem[Davis et al.(2000)]{2000MNRAS.318..387D} Davis, J., Tango, W.~J., \& Booth, A.~J.\ 2000, \mnras, 318, 387
\bibitem[de Jager \& Nieuwenhuijzen(1987)]{1987A&A...177..217D} de Jager, C., \& Nieuwenhuijzen, H.\ 1987, \aap, 177, 217
\bibitem[Duflot et al.(1957)]{1957POHP....4...11D} Duflot, M., Fehrenbach,  C., Guillaume, J., \& Ray, G.\ 1957, Publications of the Observatoire Haute-Provence, 4, 11
\bibitem[Dyck et al.(1996)]{dyc96} Dyck, H.~M., Benson, J.~A., van Belle, G.~T., \& Ridgway, S.~T.\ 1996, \aj, 111, 1705
\bibitem[Dyck et al.(1998)]{1998AJ....116..981D} Dyck, H.~M., van Belle, G.~T., \& Thompson, R.~R.\ 1998, \aj, 116, 981
\bibitem[Eddington(1921)]{ed21} Eddington, A.S., 1921, Zs. Phys., 7, 351-97
\bibitem[Eggen(1963)]{1963AJ.....68..483E} Eggen, O.~J.\ 1963, \aj, 68, 483
\bibitem[Eggen(1972)]{1972ApJ...175..787E} Eggen, O.~J.\ 1972, \apj, 175, 787
\bibitem[Famaey et al.(2005)]{2005A&A...430..165F} Famaey, B., Jorissen, A., Luri, X., Mayor, M., Udry, S., Dejonghe, H., \& Turon, C.\ 2005, \aap, 430, 165
\bibitem[Fehrenbach(1961)]{1961JO.....44..233F} Fehrenbach, C.\ 1961, Journal des Observateurs, 44, 233
\bibitem[Fernie(1983)]{1983ApJS...52....7F} Fernie, J.~D.\ 1983, \apjs, 52, 7
\bibitem[Gilmore(2007)]{2007ecf..book..205G} Gilmore, G.\ 2007, Exploring the Cosmic Frontier: Astrophysical Instruments for the 21st Century, 205
\bibitem[Gray et al.(2001)]{2001AJ....121.2148G} Gray, R.~O., Napier, M.~G., \& Winkler, L.~I.\ 2001, \aj, 121, 2148
\bibitem[Hickman et al.(1995)]{1995AJ....110.2910H} Hickman, M.~A., Sloan, G.~C., \& Canterna, R.\ 1995, \aj, 110, 2910
\bibitem[Humphreys(1970)]{1970ApJ...160.1149H} Humphreys, R.~W.\ 1970, \apj, 160, 1149
\bibitem[Johnson \& Morgan(1953)]{1953ApJ...117..313J} Johnson, H.~L., \& Morgan, W.~W.\ 1953, \apj, 117, 313
\bibitem[Johnson(1968)]{joh68} Johnson, H.~L., 1968, in Stars and Stellar Systems, 7, Nebulae and Interstellar Matter, ed. B.M. Middlehurst \& L.H. Aller (Chicago: Univ. of Chicago Press), chap. 5
\bibitem[Jura \& Kleinmann(1990)]{1990ApJS...73..769J} Jura, M., \& Kleinmann, S.~G.\ 1990, \apjs, 73, 769
\bibitem[Keenan \& McNeil(1989)]{1989ApJS...71..245K} Keenan, P.~C., \& McNeil, R.~C.\ 1989, \apjs, 71, 245
\bibitem[Keenan \& McNeil(2006)]{2006yCat.3150....0K} Keenan, P.~C., \&  McNeil, R.~C.\ 2006, VizieR Online Data Catalog, 3150, 0
\bibitem[Kharchenko(2001)]{2001KFNT...17..409K} Kharchenko, N.~V.\ 2001, Kinematika i Fizika Nebesnykh Tel, 17, 409
\bibitem[Kornilov et al.(1991)]{1991TrSht..63....4K} Kornilov, V.~G., Volkov, I.~M., Zakharov, A.~I., Kozyreva, L.~N., Kornilova, L.~N., \& et al.\ 1991, Trudy Gosudarstvennogo Astronomicheskogo Instituta, 63, 4
\bibitem[Kron(1958)]{1958PASP...70..561K} Kron, G.~E.\ 1958, \pasp, 70, 561
\bibitem[Le Sidaner \& Le Bertre(1996)]{1996A&A...314..896L} Le Sidaner, P., \& Le Bertre, T.\ 1996, \aap, 314, 896
\bibitem[Levesque et al.(2005)]{2005ApJ...628..973L} Levesque, E.~M., Massey, P., Olsen, K.~A.~G., Plez, B., Josselin, E., Maeder, A., \& Meynet, G.\ 2005, \apj, 628, 973
\bibitem[Lutz \& Kelker(1973)]{1973PASP...85..573L} Lutz, T.~E., \& Kelker, D.~H.\ 1973, \pasp, 85, 573
\bibitem[Moreno(1971)]{1971A&A....12..442M} Moreno, H.\ 1971, \aap, 12, 442
\bibitem[Morgan \& Keenan(1973)]{1973ARA&A..11...29M} Morgan, W.~W., \& Keenan, P.~C.\ 1973, \araa, 11, 29
\bibitem[Mozurkewich et al.(1991)]{1991AJ....101.2207M} Mozurkewich, D., et al.\ 1991, \aj, 101, 2207
\bibitem[Neugebauer et al.(1984)]{1984ApJ...278L...1N} Neugebauer, G., et al.\ 1984, \apjl, 278, L1
\bibitem[Oudmaijer et al.(1998)]{1998MNRAS.294L..41O} Oudmaijer, R.~D., Groenewegen, M.~A.~T., \& Schrijver, H.\ 1998, \mnras, 294, L41
\bibitem[Payne \& Chase(1927)]{1927HarCi.300....1P} Payne, C.~H., \& Chase, C.~T.\ 1927, Harvard College Observatory Circular, 300, 1
\bibitem[Perryman et al.(1997)]{1997A&A...323L..49P} Perryman, M.~A.~C., et al.\ 1997, \aap, 323, L49
\bibitem[Pickles(1998)]{pic98} Pickles, A.~J.\ 1998, \pasp, 110, 863
\bibitem[Piirola(1976)]{1976HelR....1.....P} Piirola, V.\ 1976, Observatory and Astrophysics Laboratory University of Helsinki Report, 1,
\bibitem[Price(1966)]{1966MNRAS.133..449P} Price, M.~J.\ 1966, \mnras, 133,  449
\bibitem[Richichi et al.(2005)]{2005A&A...431..773R} Richichi, A., Percheron, I., \& Khristoforova, M.\ 2005, \aap, 431, 773
\bibitem[Rufener(1976)]{1976A&AS...26..275R} Rufener, F.\ 1976, \aaps, 26, 275
\bibitem[Stefan (1879)]{Stefan1879} Stefan, J., 1879, Sitzungsberichte der mathematisch-naturwissenschaftlichen Classe der kaiserlichen Akademie der Wissenschaften, 79, 391
\bibitem[Scholz \& Takeda(1987)]{1987A&A...186..200S} Scholz, M., \& Takeda, Y.\ 1987, \aap, 186, 200
\bibitem[Shapley(1931)]{1931BHarO.881....4S} Shapley, H.\ 1931, Harvard College Observatory Bulletin, 881, 4
\bibitem[Shapley(1932)]{1932BHarO.886....9S} Shapley, H.\ 1932, Harvard College Observatory Bulletin, 886, 9
\bibitem[Thompson et al.(2002a)]{Thompson2002ApJ...577..447T} Thompson, R.~R., Creech-Eakman, M.~J., \& van Belle, G.~T.\ 2002, \apj, 577, 447
\bibitem[Thompson et al.(2002b)]{Thompson2002ApJ...570..373T} Thompson, R.~R.,Creech-Eakman, M.~J., \& Akeson, R.~L.\ 2002, \apj, 570, 373
\bibitem[Turon et al.(1993)]{1993BICDS..43....5T} Turon, C., et al.\ 1993, Bulletin d'Information du Centre de Donnees Stellaires, 43, 5
\bibitem[Unwin et al.(2008)]{2008PASP..120...38U} Unwin, S.~C., et al.\ 2008, \pasp, 120, 38
\bibitem[van Belle et al.(1999)]{1999AJ....117..521V} van Belle, G.~T., et al.\ 1999, \aj, 117, 521
\bibitem[van Belle \& van Belle(2005)]{2005PASP..117.1263V} van Belle, G.~T., \& van Belle, G.\ 2005, \pasp, 117, 1263
\bibitem[van Belle et al.(2008)]{2008ApJS..176..276V} van Belle, G.~T., et
al.\ 2008, \apjs, 176, 276
\bibitem[Yoss(1961)]{1961ApJ...134..809Y} Yoss, K.~M.\ 1961, \apj, 134, 809
\bibitem[Zdanavicius et al.(1972)]{1972VilOB..34....3Z} Zdanavicius, K., et al.\ 1972, Vilnius Astronomijos Observatorijos Biuletenis, 34, 3
\end{thebibliography}
